\documentstyle[preprint,aps,epsf]{revtex}

\draft

\preprint{\vbox{\hsize=120pt\noindent SLAC--PUB--8447 \\
September 2000 }}

\begin{document}
\pagestyle{empty}

\renewcommand{\thefootnote}{\fnsymbol{footnote}}
\title{
{\bf\large   
Measurement of $A_c$ with Charmed Mesons at SLD\footnote{Work supported by
Department of Energy contract  DE--AC03--76SF00515 (SLAC).}}
}
\author{The SLD Collaboration$^{**}$}
\address{Stanford Linear Accelerator Center\\
         Stanford University, Stanford, California, 94309\\}
\maketitle
\begin{abstract}
We present a direct measurement of the parity-violation parameter $A_c$
in the coupling of the $Z^0$ to $c$-quarks with the SLD detector.
The measurement is based on a sample of 530k hadronic $Z^0$ decays,
produced with a mean
electron-beam polarization of $|P_e| = 73 \%$.
The tagging of $c$-quark events is performed using two methods:
the exclusive reconstruction of $D^{\ast+}$, $D^+$, and $D^0$ mesons, 
and the soft-pions ($\pi_s$) produced in the decay of
$D^{\ast+}\rightarrow D^0 \pi_s^+$.
The large background from $D$ mesons produced in $B$ hadron decays
is separated efficiently from the signal using precision vertex information.
The combination of these two methods yields $A_c = 0.688 \pm 0.041.$
\end{abstract}

\vskip 0.3in
\begin{center}
{\rm Submitted to {\em Physical Review D.}}
\end{center}
\vskip 1.0in
\pagebreak

\pagestyle{plain}

\section{ Introduction }
In the Standard Model, the $Z^0$ coupling to fermions has 
both vector ($v_f$) and axial-vector ($a_f$) components. 
Measurements of fermion asymmetries at the $Z^0$ resonance probe
a combination of these components given by 
\begin{equation}
 A_f = 2v_f a_f/(v_f^2 + a_f^2). 
\label{eq:1}
\end{equation}
The parameter $A_f$ expresses the extent of parity violation 
at the $Zf\bar{f}$ vertex
and its measurement provides a sensitive test of the Standard Model.

At the Born-level, the differential cross section 
for the reaction $e^+e^- \rightarrow 
Z^0 \rightarrow f\bar{f}$ is 
\begin{equation}
 \sigma_f(z) \equiv d\sigma_f/dz 
\propto (1-A_eP_e)(1+z^2) + 2A_f(A_e - P_e) z, 
\label{eq:2}
\end{equation}
where $P_e$ is the longitudinal polarization of the electron beam 
($P_e > 0$ for net right-handed polarization)
and $z = \cos\theta$, $\theta$ being the polar angle 
of the outgoing fermion relative to the incident electron.
In the absence of electron beam polarization, the parameter $A_f$
can be extracted by isolating the term linear in $z$ via
the forward-backward asymmetry:
\begin{equation}
 A_{FB}^f(z)=\frac
{\sigma^f(z)-\sigma^f(-z)}
{\sigma^f(z)+\sigma^f(-z)}
=A_eA_f\frac{2z}{1+z^2},
\end{equation}
which also depends on the initial state electron parity-violation 
parameter $A_e$.
At the SLAC Linear Collider (SLC), the ability to manipulate the 
longitudinal polarization of the electron beam allows the isolation
of the parameter $A_f$ in Eq.~\ref{eq:2}, independent of $A_e$,
using the left-right forward-backward
asymmetry:
\begin{equation}
 \tilde{A}_{FB}^f(z)=\frac
{[\sigma_L^f(z)-\sigma_L^f(-z)]-[\sigma_R^f(z)-\sigma_R^f(-z)]}
{[\sigma_L^f(z)+\sigma_L^f(-z)]+[\sigma_R^f(z)+\sigma_R^f(-z)]}
=|P_e|A_f\frac{2z}{1+z^2},
\end{equation}
where indices $L$, $R$ refers to $Z^0 \rightarrow f\bar{f}$ decays
produced with left-handed or right-handed
polarization of the electron beam, respectively.
For a highly polarized electron beam with $|P_e|$=73\%,
$\tilde{A}_{FB}^f$ provides a
statistical advantage of $(P_e/A_e)^2$ $\sim$ 24 
in the sensitivity to $A_f$ relative to the unpolarized asymmetry.

In this paper, we present a direct measurement of the
parity-violation parameter $A_c$ for the $Zc\bar{c}$ coupling.
The $c$-quark\footnote{Throughout the paper charge-conjugate 
states are implicitly included.} 
is the only up-type quark which can be identified, 
and its measurements provides sensitive test of the standard model.
The tagging of $c$-quarks is performed using exclusively reconstructed 
$D^{\ast+}$, $D^+$, and $D^0$ mesons, as well as an 
inclusive sample of $D^{\ast+}\rightarrow D^0 \pi_s^+$ decays
identified by the soft-pion ($\pi_s$).

The charge of the primary $c$-quark is determined 
by the charge of the $D^{(\ast)}$, 
$K$ (in the $D^0$ reconstruction case),
or $\pi_s$ (in the soft-pion analysis case).
The direction of the primary quark is estimated from 
the direction of the exclusively reconstructed $D^{(*)+}$ or $D^0$ meson, 
or the jet axis in the soft-pion analysis. 
The value of $A_c$ is extracted via an unbinned maximum likelihood fit.
The fit is performed on two separate data samples:
one collected between 1993 and 1995, and the other, with an improved
vertex detector, between 1996 and 1998.
The data samples associated with these two periods comprise
150k and 380k hadronic $Z^0$ decays, respectively.

The measurement of $A_c$ presented in this paper updates and supersedes
our previous publication\cite{Ac 1993}, which was
based on a sample
of 50k hadronic $Z^0$ decays from 1993 alone.
There are several direct and indirect $A_c$ 
measurements\cite{Ac 1993,Ac SLD,Ac LEP}.
The measurement reported here is currently the most precise.

\section{ Apparatus and event selection }\label{section:2}
The measurement described here is based on 
530k hadronic $Z^0$ decays recorded in 1993-98
with the SLC Large Detector (SLD) at the SLC $e^+e^-$ collider 
at a mean center-of-mass-energy of 
91.27 GeV(1993-95) or 91.24 GeV(1996-98).
A general description of the SLD can be found elsewhere\cite{sld}.
Charged-particle tracking for the 1993-95 data sample
uses the central drift chamber (CDC)\cite{cdc} and 
VXD2\cite{vxd2} CCD pixel vertex detector. For this system,
the measured impact-parameter resolution in the transverse
(longitudinal) direction 
with respect to the beam axis can be approximated by
 $11 \bigoplus 70 / P \sin ^{5/2}\theta$ $\mu$m 
($38 \bigoplus 70 / P \sin ^{5/2}\theta$ $\mu$m), 
as a function of the track momentum $P$ (in GeV/c) 
and the polar angle $\theta$. 
In 1996, we installed the upgraded 307M pixel vertex detector
(VXD3)\cite{vxd3}, 
which provides improved impact-parameter resolution of 
 $7.8 \bigoplus 33 / P \sin ^{5/2}\theta$ $\mu$m 
($9.7 \bigoplus 33 / P \sin ^{5/2}\theta$ $\mu$m)\cite{vxd3:tracking}
in the transverse (longitudinal) direction 
with respect to the beam axis.
In addition, VXD3 extended the polar-angle coverage 
from $|\cos\theta| < 0.75$ to $|\cos\theta| < 0.85$.
Combining the CDC and the VXD, a momentum resolution of
$\sigma (P_T)/P_T = \sqrt{(.01)^2 + (.0026 P_T/GeV)^2}$
is achieved.
The Liquid Argon Calorimeter (LAC)\cite{lac} measures the energy of 
charged and neutral particles and is also used for electron 
identification. 
The barrel LAC covers the polar-angle region of $|\cos\theta| < 0.84$, and
has energy resolutions of 
$15\%/\sqrt{E\mbox{(GeV)}}$ and $65\%/\sqrt{E\mbox{(GeV)}}$ for
electromagnetic and hadronic showers, respectively.
Muon identification is provided by the Warm Iron Calorimeter
(WIC)\cite{wic}.
The \v Cerenkov Ring Imaging Detector (CRID)\cite{crid} 
provides particle identification.
In order to achieve particle identification over a wide momentum range, 
the CRID uses two different radiator systems;
liquid (C$_6$F$_{14}$) and gas (C$_5$F$_{12}$), which provide
excellent $\pi/K$ separation in the momentum range from 0.3 to 35 GeV/$c$.

The SLC operates a polarized electron beam and an unpolarized 
positron beam\cite{SLC}.
The average electron polarization measured for
the 1993-98 data sample is $|P_e| = 73 \pm 0.5\%$\cite{SLC,SLDALR}.
The SLC interaction-point (IP) size in the $xy$ plane
is 2.6 $\mu$m $\times$ 0.8 $\mu$m and its mean position is
reconstructed with a precision of $\sigma_{IP} = 4\mu$m (7$\mu$m)
using the tracks in sets of $\sim$30 sequential hadronic events
for the 1996-98 (1993-95) data sets\cite{IPmeasure}.
The event-by-event median $z$ position of tracks at their point of closest approach
to the IP in the $xy$ plane determines the $z$ position of the $Z^0$
primary vertex (PV) with a precision of $\sim$15 $\mu$m 
(35 $\mu$m) for the 1996-98 (1993-95) data.

Hadronic events are selected
by requiring at least 5 charged tracks, 
a total charged energy of at least 20 GeV/c, and
a thrust axis calculated from charged tracks satisfying
$|\cos\theta_{thrust}| < 0.87 $ (0.8 for the 1993-95 data).
In the event selection and charm reconstruction, we use the
quality tracks which satisfy
the following criteria for the 1996-98 (1993-95) samples:
\begin{enumerate}
\item At least 23 (30 for the 1993-95 data) 
      associated CDC hits; 
\item A radius of the innermost CDC hit of the reconstructed track 
   within 50 cm (39 cm) of the IP;
\item An $xy$ and $rz$ impact parameter with respect to the IP
      of less than 5cm (10cm);
\item A reconstructed polar angle $\theta$ within
    $\mid \cos \theta \mid < 0.87$ (0.80); and
\item A momentum component transverse to the beam axis greater than 0.15 GeV/c.
\end{enumerate}

As $Z^0 \rightarrow b\bar{b}$ events are also a copious source of
$D$ mesons, they represent a potential background.
We reject these events using the invariant mass of the 
charged tracks associated with the
reconstructed secondary decay vertices\cite{masstag}.
In particular, we require that there must be no vertex 
with invariant mass greater than 2.0 GeV/c$^2$.
Monte Carlo (MC) simulations indicate that this cut rejects 57\% of 
$b\bar{b}$ events while preserving 99\% of $c\bar{c}$ events.

\section{ $A_c$ measurement with exclusive charmed-meson
reconstruction}
In this analysis, we reconstruct three different $D^{(\ast)}$ meson states
for $c$-quark tagging:
the pseudo-scalar mesons $D^+$ and $D^0$, and the
vector meson $D^{\ast +}$.
This section describes the procedure for their reconstruction,
as well as the corresponding $A_c$ measurement and a discussion
of associated systematic errors.

\subsection{$D^{\ast+}$ selection}
$D^{\ast+}$ mesons are identified 
via the decay $D^{\ast+} \rightarrow D^0 \pi_s^+ $ 
followed by: 
\[\begin{array}{ll} 
 D^0 \rightarrow K^- \pi^+  & K\pi,\\
 D^0 \rightarrow K^- \pi^+ \pi^0 & Satellite,\\
 D^0 \rightarrow K^- \pi^+ \pi^- \pi^+ & K\pi\pi\pi, \mbox{ or} \\
 D^0 \rightarrow K^- l^+ \nu_l \hskip 0.5cm \mbox{($l$=e or $\mu$)} & Semileptonic.
\end{array}\]
In these decays, the charge of the underlying $c$ quark is specified
by the charge of the ``soft pion'' $\pi_s$.
No attempt is made to reconstruct the $\pi^0$ in the satellite mode,
nor to estimate
the neutrino direction or energy in the semileptonic mode.

We search for $D^{*+}$ mesons in each of the two event hemispheres,
defined by the plane perpendicular to the thrust axis, using all
quality tracks
with at least one hit in the VXD.
In the $K\pi\pi\pi$ mode, we only use tracks which have 
momentum greater than 0.75 GeV/c.
We first construct $D^0$ candidates using all combinations of
tracks corresponding to the charged multiplicity in each $D^0$ 
decay mode, with zero net charge. Here one of them is assigned the
charged kaon mass and the other(s) are assigned the charged pion mass.
In the semileptonic mode, we combine an identified electron or muon
track with another track which has opposite charge and assume 
the track to be a kaon.
Electrons are identified based on the momentum measured with the CDC
and the energy deposited in the calorimeter\cite{e+mu:ID}. Electrons
from $\gamma$-conversions are rejected. Muon candidates are identified
by the association of extrapolated CDC tracks with hits in the 
WIC\cite{e+mu:ID}.

A vertex fit is performed on the tracks forming a $D^0$ candidate, and 
we require that its $\chi^2$ probability be greater than 1\%.
The invariant mass $M$ of the $D^0$ candidates is required to
lie within the following ranges:
\[\begin{array}{ll} 
1.765 \mbox{GeV/c}^2 < M_{D^0} < 1.965 \mbox{GeV/c}^2 & (K\pi),\\
1.500 \mbox{GeV/c}^2 < M_{D^0} < 1.600 \mbox{GeV/c}^2 & (Satellite),\\
1.795 \mbox{GeV/c}^2 < M_{D^0} < 1.935 \mbox{GeV/c}^2 & (K\pi\pi\pi),\\
1.100 \mbox{GeV/c}^2 < M_{D^0} < 1.800 \mbox{GeV/c}^2 & (Semileptonic).
\end{array}\]
These reconstructed pseudo-scalar meson candidates are then combined
with a soft-pion candidate track with charge
opposite to that of the kaon candidate, thus forming the
$D^{\ast+}$ candidate.

To reconstruct the $D^{\ast+}$, we use two sets of selection criteria.
One is based on event kinematics and the other on event topology. 
The former relies on the fact that $D^{\ast+}$ 
mesons in $c\bar{c}$ events have much higher 
$x_{D^{\ast}}\equiv 2E_{D^{\ast}}/E_{CM}$, where $E_{D^{\ast}}$ is the
$D^{\ast+}$ energy, than those in $b\bar{b}$ events or random 
combinatoric background (RCBG).
The latter relies on the fact that 
$D^0$'s in $c\bar{c}$ events have a longer
3D decay length ($\sim$ 1 mm) than that for RCBG,
and originate at the primary vertex,
in contrast to those $D^0$'s in $b\bar{b}$ events
emerging from $B$ decay vertices.
We select the combinations which satisfy either condition.

In the selection based on the event kinematics, 
we require the candidate to have $x_{D^{\ast}}$ greater than $0.4$
($K\pi$, $Satellite$, and $Semileptonic$) or $0.6$ ($K\pi\pi\pi$).
For a true $D^0$ candidate, the distribution of $\cos\theta^{\ast}$,
where $\theta^{\ast}$ is the opening
angle between the direction of the $D^0$ in the laboratory frame and 
the kaon in the $D^0$ rest frame, 
is expected to be flat.
Since background events peak at $\cos\theta^{\ast} = \pm 1$, 
they are further reduced by requiring 
$|\cos\theta^{\ast}| \leq 0.9$ ($K\pi$, $Satellite$, and $Semileptonic$)
or 0.8 ($K\pi\pi\pi$).
We also require the soft-pion candidate to have momentum
greater than 1 GeV/c.
In the satellite mode,  we apply a 3D decay-length cut of
$L/\sigma_L > 1.5$ on the reconstructed $D^0$ vertices to reduce
the RCBG. 
(The average decay-length resolution is $<\sigma_L>$ 
$\sim$ 150 $\mu$m.)

In the selection based on the event topologies, 
we require the reconstructed  $D^0$ vertices to have 
3D decay-length significance $L/\sigma_L > 2.5$, and
the 
$xy$ impact parameter of the $D^0$ momentum
vector to the IP to be less than 20 $\mu$m ($K\pi$ and $K\pi\pi\pi$) or
30 $\mu$m ($Satellite$ and $Semileptonic$).
The latter cut is effective in rejecting $D$ decays in $b\bar{b}$ events.
Since these $D$'s have significant $P_T$ relative to the parent
$B$ flight direction, and the $B$'s themselves have a significant
flight length ($\sim$ 3.5 mm),
many of these $D$'s do not appear to originate from the primary vertex.
A cut of $x_{D^{\ast}}$ greater than $0.3$ ($K\pi$,
$Satellite$, and $Semileptonic$) 
or 0.4 ($K\pi\pi\pi$) is also applied.
Fig.~1 shows the distribution of 
$xy$ impact parameter of the $D^0$ relative to the IP
for the decay of
$D^{\ast+}\rightarrow D^0 \pi_s^+$, $D^0 \rightarrow K^- \pi^+$. 
In this figure, we do not reject $B$-decay candidate events with 
the invariant mass cut of the reconstructed secondary 
vertices 
described above, 
only for the purpose of showing how the 
$xy$-impact-parameter cut is
effective in rejecting the $B$-decay background.
After applying the invariant mass cut of the reconstructed secondary 
vertices, 
34\% of the remaining $B$-decay background events are rejected by 
the $xy$-impact-parameter cut.

The overlaps of the sets of candidates from the event kinematics
and topology analysis are 53\% ($K\pi$), 50\% (satellite), 28\%
($K\pi\pi\pi$), and 36\% (semileptonic).
In the $K\pi\pi\pi$ sample, there may be multiple 
$D^0$ candidates in a single event which pass the above cuts.
To avoid double counting and to reduce the background, 
we select the $D^0$ candidate with the lowest vertex $\chi^2$.

Having selected a candidate, we form
the mass difference $\Delta M = M_{D^{\ast}} - M_{D^0}$.
The mass difference spectra for the four reconstructed $D^{\ast +}$
decay modes are shown in Fig.~2. 
For all decay modes, clear peaks around $\Delta M = 0.14$ GeV/c$^2$ appear
due to the $D^{\ast+}$ to $D^0$ transition.
We include the candidates in the signal sample provided
$\Delta M$ is less than 0.148 \mbox{GeV/c}$^2$ ($K\pi$ and $K\pi\pi\pi$),
less than 0.155 \mbox{GeV/c}$^2$
($Satellite$), and less than 0.16 \mbox{GeV/c}$^2$ ($Semileptonic$).
The side-band region is defined as 0.16 $< \Delta M <$ 0.20 \mbox{GeV/c}$^2$
(0.17 $< \Delta M <$ 0.20 \mbox{GeV/c}$^2$ for the $Semileptonic$ mode),
and is used to estimate the RCBG
contamination in the signal region.
In the figure, the MC predictions 
for the reconstructed $D^{\ast+}$
(open) and RCBG (hatched) are also presented.
For the MC prediction, the relative normalizations of signal and RCBG
shapes are adjusted so that the predicted numbers of events match
those observed in the data signal and side-band regions.
Averaged over the various modes, this procedure requires adding 
10\% to the MC signal and 5\% to the MC RCBG.
The number of the selected candidates as well as the
contributions of $c, b \rightarrow D$ and RCBG estimated by MC
are summarized in Table~\ref{tab:candidate}. 

\subsection{$D^+$ and $D^0$ selection}
The $D^+$ and $D^0$ mesons are identified via the decay channels
\[\begin{array}{l}
D^+ \rightarrow K^- \pi^+ \pi^+  \\
D^0 \rightarrow K^- \pi^+.
\end{array}\]       
These modes are reconstructed by considering all quality tracks in each
hemisphere which have VXD hits.
In the $D^+$ reconstruction, we additionally require each track to have
a momentum of greater than 1 GeV/c.

For the $D^+$ reconstruction, we combine two same-sign tracks, 
assumed to be pions, with an opposite-sign track, 
assumed to be a kaon.
We require that  $x_{D^+}$ be greater than 0.4, and
$\cos\theta^{\ast}$ be greater than -0.8,
where $\theta^{\ast}$ is the opening
angle between the direction of the $D^+$ in the laboratory frame and 
the kaon in the $D^+$ rest frame.
To reject $D^{\ast+}$ decays, the differences between
$M_{K^-\pi^+\pi^+}$ and $M_{K^-\pi^+}$ are formed for each of the
pions, and both are required to be greater than 0.16 GeV/c$^2$.
To remove RCBG, 
we require that the $\chi^2$ probability of the good vertex fit be 
greater than 1$\%$, and that the 3D decay-length significance 
$L/\sigma_L$ be greater than 3.0.
To reject $D^+$'s from $b\bar{b}$ events,
the angle between the $D^+$ momentum vector and
the vertex flight direction is required 
to be less than 5 mrad in 
$xy$ and less than 20 mrad in $rz$.
Here we use the angular information instead of the 
impact-parameter information. 
We can strongly constrain the $D^+$ to originate from the IP
with the angular information, because of its large decay length.

To form the $D^0$ vertices, tracks identified as charged kaons,
by the requirement that the CRID log-likelihood\cite{K:ID} for the 
$K$ hypothesis exceeds that for the $\pi$ hypothesis by at 
least 3 units,
 are combined with an opposite-charge track, assumed to be a pion.
We use the CRID information for this mode only.
To reject background we require $x_{D^0}$ be greater than $0.4$.
We require that the vertex fit have $\chi^2$ probability greater than
1$\%$ and the 3D decay-length cut 
$L/\sigma_L$ be greater than 3.0.
To reject the $D^0$'s from $D^{\ast+}$ decays, the differences between
$M_{K^-\pi^+\pi^+}$ or $M_{K^-\pi^+\pi^-}$, 
and $M_{K^-\pi^+}$ are formed for all other tracks
in the same hemisphere,
and these are required to be greater than 0.16 GeV/c$^2$.
Finally, 
to reject $D^0$'s from $b\bar{b}$ events,
we require that the 
$xy$ impact parameter of the $D^0$ momentum
vector relative to the IP be less than 20 $\mu$m.

$D^+$ and $D^0$ candidates in the ranges of
1.800 $< M_{K^-\pi^+\pi^+} <$ 1.940 GeV/c$^2$ and 
1.765 $< M_{K^-\pi^+} <$ 1.965 GeV/c$^2$, respectively, are regarded as signal.
The side-band regions are defined as 
1.640 $< M_{K^-\pi^+\pi^+} <$ 1.740 \mbox{GeV/c}$^2$ and 
2.000 $< M_{K^-\pi^+\pi^+} <$ 2.100 \mbox{GeV/c}$^2$ for $D^+$, and
2.100 $< M_{K^-\pi^+} <$ 2.500 \mbox{GeV/c}$^2$ for $D^0$.
In Fig.~3, the invariant mass spectra for
the resulting $D^+$ and $D^0$ signals are plotted.
The backgrounds in the signal regions are estimated from the MC
in the same manner as in the $D^{\ast +}$ analysis. 

\subsection{ Measurement of $A_c$ }\label{section:3.3}
Using the six decay modes, 
we select 3967 $D^{\ast +}, D^+$, and  $D^0$ candidates 
from 1993-98 SLD data.
The estimated composition is 2829$\pm$35  $c \rightarrow D$ signal, 
281$\pm$11 $b \rightarrow D$, and 857$\pm$19 RCBG.
These $c \rightarrow D$ signals correspond to a selection efficiency 
for $c\bar{c}$ events of 3.9\%.
The results for the number of selected candidates are summarized in
Table~\ref{tab:candidate}. 

The charge of the primary $c$-quark is determined
by the charge of the $D^{(\ast)}$, or  $K$ (in the $D^0$ case).
The direction of the primary quark is estimated from 
the direction of the reconstructed $D$ meson.
Fig.~4 shows $q \cos\theta_D$ distributions,  
for the selected $D$ meson sample separately for left- and right-handed
electron beams. 
Here, $q$ is the sign of the charge of the primary $c$-quark
and $\theta_D$ is the polar angle of the reconstructed
$D$ meson.  

To extract $A_c$, we use an unbinned maximum likelihood fit based on
the Born-level cross section for fermion production in $Z^0$-boson
decay. 
The likelihood function used in this analysis is
\begin{eqnarray}
\ln{\cal L}= \sum^{n}_{i=1} 
& \ln & \{
P_c^j(x_D^i) \cdot [(1-P_eA_e)(1+y_i^2)
+2(A_e-P_e)y_i \cdot A_c^D 
] \nonumber \\
 & + &  
P_b^j(x_D^i) \cdot [(1-P_eA_e)(1+y_i^2)
+2(A_e-P_e)y_i\cdot A_b^D 
] \nonumber \\
 & + &  
P_{RCBG}^j(x_D^i) \cdot [(1+y_i^2)
+2A_{RCBG}y_i]
\}
\label{eq:likelihood}
\end{eqnarray}
where $y = q cos\theta_D$,
$n$ is the total number of candidates, and
the index $j$ indicates each of the six charm decay modes.

$A_c^D$ and $A_b^D$ are the asymmetries from $D^{\ast+}$,
$D^+$, and $D^0$ mesons in $c\bar{c}$ and $b\bar{b}$ events, 
respectively. 
We treat $A_c^D$ as a free parameter, while
$A_b^D$ is fixed.
$A_b^D$ is estimated in a similar manner to Ref.\cite{A_bD}.
We start with the Standard Model prediction\cite{SM}, $A_b = 0.935$,
and assign it an error of $\pm0.025$ from the average value of
SLD measurements of 0.911 $\pm$ 0.025\cite{LEP-SLD-WG00}.
This $b$-quark asymmetry is diluted by $B^0$-$\bar{B^0}$ mixing and 
the wrong-sign $D$ meson from the $W^-$ in $b \rightarrow cW^-$, 
$W^- \rightarrow \bar{c}s$ decay.
The effective $b$ asymmetry can be expressed by correcting 
with two dilution factors:
\begin{equation}
 A_b^D = A_b \times 
( 1 - 2\chi_{mixing})( 1 - 2\chi_{W^-\rightarrow\bar{c}s}). 
\label{eq:ab}
\end{equation}
The value of $\chi_{mixing}$ is deduced from the $D$-meson production
rates through $B$ decays.  
We estimate the $B \rightarrow D$ source fractions from MC.
Using the fractions and the $\chi$ values of 
$\bar{\chi} = 0.1186 \pm 0.0043$\cite{LEP-SLD-WG00}
and $\chi_{d} = 0.156 \pm 0.024$\cite{mixing-PDG},
we derive the $\chi_{mixing}$ value for 
$D^{\ast +}$, $D^+$, or $D^0$.
The value of $\chi_{W^-\rightarrow \bar{c}s}$, 
the correction for wrong-sign $D$ mesons from the $W^-$ in 
$b\rightarrow cW^- $ decay,  is also estimated from MC.
We obtain $\chi_{W^-\rightarrow \bar{c}s} = 0.023 \pm 0.006$
for the average of $D^{\ast+}$, $D^+$, and $D^0$ mesons, and 
$0.021 \pm 0.006$ for $D^{\ast+}$ mesons only.
Here the errors include the theoretical error of 30\% 
coming from $Br(b \rightarrow c\bar{c}s) = 22 \pm 6\%$\cite{ncc}.
The former and latter $\chi_{W^-\rightarrow \bar{c}s}$ values 
are used for exclusive $D$ reconstruction and inclusive soft-pion 
analysis, respectively.
By combining these two dilutions, we obtain
\[\begin{array}{llll} 
A_b^D & = & 0.657 \pm 0.025 & \mbox{for } D^{\ast+},\\
      & = & 0.655 \pm 0.026 & \mbox{for } D^+ \mbox{ and},\\
      & = & 0.762 \pm 0.023 & \mbox{for } D^0.
\end{array}\]

To check the $A_b^D$ value, we measure $A_b^D$ for $D^{\ast+}$
using the 1996-98 experimental data.
In this measurement, we select $D^{\ast+}$ mesons in the decay, 
$D^{\ast+} \rightarrow D^0 \pi_s^+ $ 
followed by 
 $D^0 \rightarrow K^- \pi^+$,   
 $D^0 \rightarrow K^- \pi^+ \pi^0$, or 
 $D^0 \rightarrow K^- \pi^+ \pi^- \pi^+$.
The $b\bar{b}$ events are selected by requiring 
that the invariant mass for the reconstructed secondary vertices be 
greater than 2 GeV/c$^2$ for at least one of the two event hemispheres.
In order to select the $D^{\ast+}$ mesons, 
we apply similar cuts to those used
to select the $D^{\ast+}$ mesons from $c$-quarks, 
but without any $xy$ impact parameter cut to reject $D^{\ast+}$'s 
from $b$-quarks.
We select 2196 $D^{\ast}$ candidates with the fractions 
of 63\% $b \rightarrow D$, 2\% $c \rightarrow D$, and 
35\% RCBG.
Using this sample, we measure $A_b^D = 0.58 \pm 0.10$, 
which is consistent with our assumed $A_b^D$ value for $D^{\ast+}$.
The error of 0.10 is treated as a systematic error of $A_b^D$.

We also check the effect of the decay-length cut of the reconstructed
$D$ mesons. In this analysis, we apply the decay-length cut of
$L/\sigma_L > 1.5 \sim 3.0$ (depending on the charm decay mode)
to reject RCBG. This cut may increase the effective value of 
$\chi_{mixing}$. Using our MC, we estimate the effect 
of this cut to be small ($\Delta\chi_{mixing}/\chi_{mixing}=3\%$).
 
$A_{RCBG}$ is the analog of $A_c$ for the RCBG, 
and we expect it to be very small.
The asymmetry in the side-band region is measured as
$-0.0006 \pm 0.0031$, 
and is assumed to be zero.
For $A_e$, we have taken 
$A_e$ = 0.1513 $\pm$ 0.0022 from  the SLD measurement\cite{SLDALR}.

$P_c^j$, $P_b^j$,  and $P_{RCBG}^j$ are the probabilities 
that a candidate from the $j$th decay mode 
is a signal from $c\bar{c}$, $b\bar{b}$,
or RCBG.
The determination of these functions is based on the relative fractions
and the $x_D$ distributions for the six decay modes. 
They are defined as:

\begin{eqnarray}
P_c(x_D) &=& \frac{N_{signal}(x_D)}{N_{total}(x_D)}\cdot
           \frac{f_c(x_D)}
                {f_c(x_D) + f_b(x_D)} 
\nonumber \\
P_b(x_D) &=& \frac{N_{signal}(x_D)}{N_{total}(x_D)}\cdot
           \frac{f_b(x_D)}
                {f_c(x_D) + f_b(x_D)}
\label{eq:probs} 
\\
P_{RCBG}(x_D) &=& \frac{N_{BG}(x_D)}{N_{total}(x_D)}; 
\nonumber 
\end{eqnarray}
where $N_{total}(x_D)$ is the observed number of $D$ mesons, 
and $N_{BG}(x_D)$ is that of background events, in the $x_D$ bin.
Using the $x_D$ distributions for the reconstructed $D$ mesons and 
side-band events, we determine the ratio 
$N_{BG}/N_{total}$ in each $x_D$ bin.
The ratio $N_{signal}/N_{total}$ is given by the relation 
$N_{signal}/N_{total} = 1 - N_{BG}/N_{total}$ in each bin.
Figs.~5~(a)-(f) show the $x_D$ distributions for six decay
modes, which are used in this determination.

The functions $f_{c}(x_D)$ and $f_b(x_D)$ describe the fraction of 
$D$ mesons in the $c$ and $b$ decays, respectively, 
and are expressed as 
\begin{equation}
f_{c(b)} = \omega_{c(b)}\cdot d_{c(b)}(x_D),
\label{eq:ctoD}
\end{equation} 
where $d_{c(b)}(x_D)$ describes the shape of $x_D$ 
distributions in $c(b) \rightarrow D$, and 
$\omega_{c(b)}$ represents the total fraction of the
$c(b) \rightarrow D$ for the reconstructed $D$ candidates. 
We obtain the function $d_{c(b)}(x_D)$ from MC, and 
the values of $\omega_c$ and $\omega_b$ are 
derived from Table~\ref{tab:candidate}.
The ratio $f_{c(b)}/(f_c + f_b)$ gives the probability that
a $D$ candidate is from a primary $c$($b$) quark.

Performing the maximum likelihood fit to the data sample, we measure
$A_c = 0.671 \pm 0.096$ (1993-95) and 
$A_c = 0.681 \pm 0.047$ (1996-98).
As a check, we also determine $A_c$ with 
a simple binned fit of the type described in Ref.\cite{e+mu:ID}. We find
$A_c = 0.731 \pm 0.102$ (1993-95) and 
$A_c = 0.666 \pm 0.049$ (1996-98); 
which are consistent with the values above.

\subsection{ QCD and QED correction }\label{section:3D}
%
As a result of hard gluon radiation, the extracted value of
$A_{c(b)}$ is somewhat different than its Born-level value 
in Eq.~\ref{eq:1}.
To account for this,
the fit parameter $A_{c(b)}$ in the
likelihood function is replaced with the first-order
corrected parameter $A_{c(b)}(1 - \Delta^{c(b)}_{QCD}(\cos\theta))$
with $\Delta^{c(b)}_{QCD}(\cos\theta) = C_{c(b)}
\Delta^{c(b)}_{QCD,SO}(\cos\theta)$,
where $\Delta^{c(b)}_{QCD}$ indicates the magnitude of 
the leading-order (LO) QCD correction for $c$($b$)-quark production, and
$\Delta^{c(b)}_{QCD,SO}$ is the LO QCD correction calculated
by Stav and Olsen including the quark-mass effect\cite{stav&olsen}.
The factor $C_{c(b)}$ takes into account the mitigation of the effects of gluon
radiation due to the analysis procedure.
For example, the requirement that $D$ mesons have high $x_D$ values
selects against events containing hard gluon radiation, 
reducing the overall effect of gluon radiation 
on the observed asymmetry.

The correction factor $C_{c(b)}$ is estimated with the MC
by comparing the effects of QCD radiation, for the  JETSET
Parton Shower model, with and without the full analysis including
detector simulation:
\begin{equation}
C_q = {A_{q {\overline q}}^{gen} - A_{PS}^{meas} \over
       A_{q {\overline q}}^{gen} - A_{PS}^{gen} } \,(q = c,b), 
\label{eq:8}
\end{equation}
where the superscripts `$gen$' and `$meas$' refer to
the MC asymmetries for generator level (Parton Shower Model simulation only)
and fully analyzed events, respectively.
These MC asymmetries are determined by doing
a fit to the form
\begin{equation}
  A {2 \cos\theta \over 1 + \cos^2\theta}
\end{equation}
in bins of $\cos\theta$. We obtain
$C_c = 0.27 \pm 0.10$ and 
$C_b = 0.17 \pm 0.08$ for $c$-quark and $b$-quark, respectively.
Applying the first-order QCD correction with the 
correction factors $C_{c(b)}$, leads to a 1.0\% increase of 
$A_c$.

In this analysis, we have also considered the effects of
next-to-leading order (NLO) gluon radiation.
The NLO QCD correction is written as:
\begin{equation}
 \Delta_c^{O(\alpha_s^2)} = \left( \frac{\alpha_s}{\pi} \right)^2
\times 4.4 \times C_c + \Delta_{gs}.
\end{equation}
where the first term is from hard gluon emission\cite{Hard gluon radiation}.
We use the same correction factor $C_{c}$ 
as in Eq.~\ref{eq:8}.
The second term $\Delta_{gs}$ accounts for the effects of the process
 $g\rightarrow c\bar{c}$ for gluons
which arise during the shower and fragmentation processes.

The effects of gluon splitting have been taken into account
by analyzing the MC as if it were data, with and
without events with gluon splitting. The resulting difference
must be scaled to account for the difference between the JETSET
gluon splitting rates and the currently measured values for these
rates. The rate for gluon splitting to charm quark pairs in
JETSET is 0.0136 per hadronic event, and 
the current LEP average\cite{LEP-SLD-WG00} is 0.0319 $\pm$ 0.0046, 
yielding a scale factor of $2.35 \pm 0.34$. 

The second-order QCD correction increases $A_c$ by 0.4\%.
Applying the first- and second-order QCD corrections, we obtain 
$A_c = 0.681 \pm 0.097 $ (1993-95) and 
$A_c = 0.690 \pm 0.047 $ (1996-98).

Using ZFITTER(6.23)\cite{SM}, we estimate QED corrections including
initial- and final-state radiation, vertex correction, $\gamma$
exchange, and $\gamma$-$Z$ interference. We use the input values
$m_{top}=175$ GeV/c$^2$ and $m_{Higgs}=150$ GeV/c$^2$.
These corrections increase $A_c$ by 0.2\%.
Applying the QED corrections, we obtain 
$A_c = 0.682 \pm 0.097 $ (1993-95) and 
$A_c = 0.691 \pm 0.047 $ (1996-98).

\subsection{ Systematic errors }
The following systematic errors have been estimated and are summarized in
Table~\ref{tab:systematic_ex}:
\begin{itemize}

\item
The largest uncertainties are due to the RCBG, 
arising from the statistics of the MC 
and side-band events, which are used 
to determine the fraction of the RCBG in the signal,
and the shape of RCBG $x_D$ distribution which is 
determined by side-band events.
The uncertainty of the RCBG $x_D$ shape is estimated
by comparing the $x_D$ distributions for MC RCBG events
and for side-band events.

\item
There is a difference in acceptance between signal and RCBG event samples.
In this analysis, we determine the RCBG probability function 
as a function of $x_D$. 
This is correct 
if the ratio between the signal and 
RCBG acceptance is constant over the different 
$\cos\theta$ regions. 
In order to study this, we compare the RCBG $|\cos\theta|$ 
distribution obtained from the side-band region and 
that from the signal region events weighted by the 
RCBG probability function $P_{RCBG}(x_D)$ in Eq.~\ref{eq:likelihood}.
These two distributions become significantly different starting at 
$|\cos\theta| \sim 0.65$. Hence,
we apply an acceptance cut of $|\cos\theta_D|<0.65$,
then regard the difference between with and without the 
cut as a systematic uncertainty.

\item
We expect the asymmetry of RCBG to be very small, and 
take a central value of $A_{RCBG}=0$. 
Since the asymmetry of the side-band events is
measured to be $-0.0006 \pm 0.0031$,
we take -0.0037 as a lower limit on $A_{RCBG}$.

\item
We vary $f_{b \rightarrow D}/(f_{b\rightarrow D} + f_{c\rightarrow
D})$, the fraction of $D$ mesons from $Z^0 \rightarrow b\bar{b}$, 
by $\pm$20\% to account for differences between our MC and the range
of measurements of $D^{(\ast)+}$ production 
in $Z^0$ decay\cite{A_bD,Dmeasure}.

\item
The effect of the uncertainty of $A_b^D$ is estimated by varying 
$\delta A_b^D = \pm 0.10$, where the error is from the statistical 
error of our $A_b^D$ measurement by using experimental data.
In Table~\ref{tab:systematic_ex},
we show the resultant error in $A_c$ coming from 
the uncertainty in $A_b$ ($0.935 \pm 0.025$) separately
from the uncertainty in the mixing parameter.

\item
The systematic error on the fragmentation function is estimated
by modifying the $x_D$ distributions in heavy-quark fragmentation.
In our MC sample, we use Peterson fragmentation 
and the average $x_D$ values are $\langle x_D \rangle =$ 0.508 and 
0.318 for $c \rightarrow D$ and $b \rightarrow D$, respectively.   
We change the values by 
$\Delta \langle x_D \rangle = \pm0.015 (\pm0.010)$ for $c(b) \rightarrow D$.

\item
Our sensitivity to the RCBG $x_D$ distribution is checked by 
performing the analysis with $P_{RCBG}$ derived from the MC background
instead of the data side-bands.

\item
The shapes of the $x_D$ distributions 
in $c(b) \rightarrow D$, expressed as $d_{c(b)}(x_D)$ 
in Eq.~\ref{eq:ctoD}, are obtained by
fitting to the MC $x_D$ distributions.
The sensitivity to this procedure
is checked by performing the analysis with 
a binned MC $x_D$ distribution.

\item 
We assume
$A_e$ = 0.1513 $\pm$ 0.0022, and estimate
this systematic error by varying $A_e$ within the error. 
The precision of the polarization measurements are $\Delta P_e =$
1.1\% (1993), 0.5\% (1994-95), and 0.4\%(1996-98)\cite{SLC,SLDALR}.
We estimate the systematic error due to polarization uncertainties 
by varying $P_e$ with these errors.

\item
We consider two sources of uncertainties on the leading  
order QCD correction: The uncertainty on $\alpha_{s}$ and 
the uncertainty in the estimation of the correction factor
due to the analysis bias.
The range of $\alpha_s$ chosen for the analysis is $0.118 \pm 0.007$,
while that for the correction factor is 
$0.27\pm 0.10$ for $c$-quark or $0.17\pm 0.08$ for $b$-quark, 
as described in Section~\ref{section:3D}.

\item
In order to estimate the hard-gluon-radiation uncertainty in the 
second-order QCD correction,
we vary the magnitude of the correction by 50\% of itself.
We use the experimental error for the uncertainty in 
gluon splitting into $c\bar{c}$.

\end{itemize}

The total systematic errors are 0.034 and 0.021 for
1993-95 and 1996-98 SLD runs, respectively.

\subsection{ Results }
We obtain the following results for the measurements using 
exclusive channels:
$A_c = 0.682 \pm 0.097(stat.) \pm 0.034(sys.)$ (1993-95) and 
$A_c = 0.691 \pm 0.047(stat.) \pm 0.021(sys.)$ (1996-98).
The combined result is:
$$A_c = 0.690 \pm 0.042(stat.) \pm 0.021(sys.)$$

\section{ Inclusive soft-pion analysis}
In this analysis, $c$-quarks are identified by the presence of 
soft pions from the decay $D^{\ast+}\rightarrow D^0 \pi_s^+$.
Since this decay has a small
Q value of $m_{D^\ast} - m_{D^0} - m_{\pi}$ = 6 MeV$/c^2$,
the maximum transverse momentum of the $\pi_s$ with respect to the 
$D^{\ast +}$ flight direction is only 40 MeV/c.

\subsection{ Jet reconstruction and soft-pion selection}
We select hadron events and reject $b\bar{b}$ events
by using the same criteria described in Section~\ref{section:2}. 
The $D^{\ast+}$ flight direction is approximated by the jet direction,
where charged tracks and neutral clusters are
clustered into jets, using an invariant-mass (JADE) 
algorithm.
In the jet clustering, particles are merged together in
an iterative way if their invariant mass is less than 4.6 GeV/c$^2$.
We only use the tracks and clusters 
which have the momentum of greater than 1.2 GeV/c and 1.0 GeV/c,
respectively, to form the jet.
The tracks are required to satisfy the track quality cuts described in
Section~\ref{section:2} and to have vertex hits.

The jets must satisfy the following criteria:
\begin{enumerate}
\item At least 3 charged tracks;
\item At least one track with momentum P $>$ 5 GeV/c; 
\item The net charge of the jet, $\Sigma q$, should be $|\Sigma q|$ $\le$ 2;
\item Sum of the largest and second largest 3D normalized impact parameters 
of the tracks $>$ 2.5 $\sigma$; and
\item There is at least one opposite-charged-track pair which has
   $\chi^2$ probability of two tracks coming from the same
   vertex greater than 1\%.
\end{enumerate}
The criteria 2) and 3) are effective to reduce the huge RCBG.
The criterion 4) rejects the light flavor events.
The criterion 5) relies on the fact
that it is likely that the $D^0$ decays into at least 
one pair of oppositely charged tracks.

After selecting the jet candidates, 
we look for the soft-pions 
using a momentum cut of $1 < P < 3$ GeV/c and an impact-parameter 
cut of less than $2\sigma$ from the IP. 
Since soft-pions in $c\bar{c}$ events have much higher momentum 
than those in $b\bar{b}$ events, the former criterion rejects 
such soft-pions from $b\bar{b}$ events. 
The latter criterion is also effective to reduce the 
soft-pions from $b\bar{b}$, 
because $D^{\ast}$ decays from $b\bar{b}$ events have significant 
transverse momentum relative to the parent $B$ flight direction, 
and they do not appear to originate from the primary vertex
due to the $B$ lifetime.

Using the selected soft-pion candidates, 
the momenta transverse to the jet axis. $P_T$, are calculated.
Fig.~6~(a) shows the $P_T^2$ distribution for 
the soft-pion candidate tracks. 
The peak around $P_T^2=0$ is from 
charm signal.
We define $P_T^2 < 0.01$ (GeV/c)$^2$ as the
signal region, where a signal-to-background ratio of 1:2 is observed.
From 1993-98 data, 12992 soft-pion candidates are selected in the region.

\subsection{ BG determination and $A_c$ measurement}
To evaluate the number of the $D^{\ast+}\rightarrow D^0 \pi_s^+$ decays, 
a fit to the observed $P_T^2$ distribution is performed using 
the signal plus background shape.
The signal shape is assumed to be a simple exponential
$$ S(P_T^2) = \alpha \exp(-P_T^2/\beta). $$
We obtain $\beta = 0.00471 \pm 0.00007$ 
by fitting the MC spectrum of $D^{\ast+}\rightarrow D^0 \pi_s^+$ 
decays and fix the value of $\beta$ to fit the experimental data.
For the background shape, we try two kinds of functions
with three free parameters each:
\begin{eqnarray}
 F_1(P_T^2) &=& a / (1 + bP_T^2 + c(P_T^2)^2), \nonumber \\ 
 F_2(P_T^2) &=& a' + b' \exp(-P_T^2/c'). \nonumber
\end{eqnarray}
The fit results are illustrated in Fig.~6~(a),
where we show the extrapolation of $F_1(P_T^2)$ (dashed line) 
and $F_2(P_T^2)$ (dotted line). 

The observed signal in 1993-98 data is
4291 $\pm$ 147 ($\chi^2/\mbox{ndf} = 219.0/196$) 
with $S(P_T^2) + F_1(P_T^2)$ and 
4032 $\pm$ 124 ($\chi^2/\mbox{ndf} = 224.0/196$) 
with $S(P_T^2) + F_2(P_T^2)$, 
where the fit is performed in each case
for $P_T^2 < 0.1$ GeV/c.
We choose $F_1(P_T^2)$ for the background shape to measure the $A_c$, 
because of its smaller $\chi^2/\mbox{ndf}$ value. 
The difference between these two functions is regarded 
as a systematic error. 

We determine the relative normalizations of signal and background 
for the MC prediction using the above fit to the data.
Fig.~6~(b) shows the detailed $P_T^2$ distribution from the 
MC prediction with this normalization. 
We also overlay the background shape extrapolated by 
the fitting with $S(P_T^2) + F_1(P_T^2)$ (dashed line).
Using the MC, we estimate the contributions of 
$c \rightarrow D^{\ast+}$ and $b \rightarrow D^{\ast+}$
as 3791$\pm$39 and 500$\pm$14, respectively, in 1993-98 data.

In order to ensure that there is 
little room for non-$D^{\ast}$ sources of slow pions 
in the data, we compared the signal obtained by fitting
to the experimental data and the number of $D^{\ast}$'s expected by MC.
Here normalization of the MC is determined by the number of hadronic 
events. Using MC, we estimate the number to be 4507$\pm$57.
Comparing this number and the obtained experimental number 
of 4291 $\pm$ 147, we conclude that other charm-decay sources
in the experimental data are small.

The direction of the primary quark is estimated from 
the jet axis, and 
the charge of the primary $c$-quark is determined 
by the charge of the $\pi_s$.
Fig.~7 shows the $q \cos\theta_D$ distributions,
where $q$ is the sign of the primary $c$-quark, and $\theta_D$
is the polar angle of the jet axis, 
for the selected $D^{\ast+}$ sample separately for left- and right-handed
electron beams.

To extract $A_c$, we use an unbinned maximum likelihood fit, 
using a likelihood function similar to the exclusive $D$ reconstruction 
analysis (Eq.~\ref{eq:likelihood}).
We regard the $A_c$ as a free parameter, and fix the asymmetry of
$D^{\ast +}$ from $b\bar{b}$ events, $A_b^D$.
This value is obtained by following the similar procedure
described in Section~\ref{section:3.3}).

We expect the asymmetry for the BG, $A_{BG}$, to be very small
and assume it to be zero.
Using the MC, we measure the asymmetry of the background 
to be $0.009 \pm 0.017$.

For the probabilities
$P_c$, $P_b$,  and $P_{RCBG}$ in Eq.~\ref{eq:likelihood}, 
we used the following functions: 
\begin{eqnarray}
P_c(P,P_T^2) &=& \frac{N_{signal}(P,P_T^2)}{N_{total}(P,P_T^2)}\cdot
           \frac{f_c(P)}
                {f_c(P) + f_b(P)}
\nonumber \\
P_b(P,P_T^2) &=& \frac{N_{signal}(P,P_T^2)}{N_{total}(P,P_T^2)}\cdot
           \frac{f_b(P)}
                {f_c(P) + f_b(P)}
\nonumber \\
P_{RCBG}(P,P_T^2) &=& \frac{N_{BG}(P,P_T^2)}{N_{total}(P,P_T^2)}; 
\label{eq:spi}
\end{eqnarray}
where $P$ and $P_T^2$ indicate the
momentum and 
the squared transverse momentum to the $D^{\ast}$ jet axis 
for soft-pion tracks, respectively.
$N_{total}$ and $N_{BG}$ are the observed number of soft-pion
candidates and that of background in each $P$ and $P_T^2$ bin, respectively.
We estimate $N_{BG}$ from MC, and the relation 
$N_{signal}/N_{total} = 1 - N_{BG}/N_{total}$
gives the ratio $N_{signal}/N_{total}$.
Fig.~8 shows the momentum distributions for experimental data 
and MC predictions. 
Figs. 6 and 8 are used for this estimation.

The function $f_{c(b)}$ in Eq.~\ref{eq:spi}
describes the fractions of $D$ mesons in the $c(b)$ decays,
and the ratio $f_{c(b)}/(f_c + f_b)$ gives the probability that
$D$ candidate is from a primary $c$($b$) quark.
We regard $f_{c(b)}$ as a function of soft-pion momentum, $P$.
The function is expressed as $f_{c(b)} = \omega_{c(b)}\cdot
d_{c(b)}(P)$. 
Here $d_c(b)$ is determined by the shape of 
MC soft-pion momentum distributions in $c(b) \rightarrow D$
and $\omega_{c(b)}$ is the estimated total fraction of the
$c(b) \rightarrow D$ among the selected candidates.

Performing the maximum likelihood fit to the data sample, we measure
$A_c = 0.654 \pm 0.125$ (1993-95) and 
$A_c = 0.673 \pm 0.056$ (1996-98).
As a check, we also measure $A_c$ with 
a simple binned fit as 
$A_c = 0.520 \pm 0.164$ (1993-95) and 
$A_c = 0.665 \pm 0.085$ (1996-98), 
which are consistent with the above values.

The first- and second-order QCD correction and QED correction
are applied with the same method as in the exclusive 
$D$ reconstruction analysis.
In the QCD correction, the correction factor due to the 
analysis bias is estimated as 
$C_c = 0.40 \pm 0.14$ for $c$-quark and
$C_b = 0.19 \pm 0.09$ for $b$-quark.
Applying the first- and second-order QCD correction 
with this factors, and QED correction, 
we obtain 
$A_c = 0.669 \pm 0.127 $ (1993-95) and 
$A_c = 0.689 \pm 0.057 $ (1996-98).

\subsection{ Systematic errors }
The estimated uncertainties in this analysis are 
summarized in Table~\ref{tab:systematic_ex}, 
where we show average systematic errors for the 1993-98 data. 
In the soft-pion analysis, we use the same procedures to 
estimate the systematic errors as those in the exclusive 
$D^{(\ast)}$ reconstruction analysis in many sources. 
Here we only explain error sources where we take a different method.

The largest uncertainties are due to the imperfect knowledge of the
background fraction and its shape.
The background is determined by fitting to the 
$P_T^2$ distribution of the experimental data, and we try
two functions $F_1$ and $F_2$ described above. 
In order to estimate the background fraction uncertainty,
we fix the background shape as $F_1$, and change its height 
so as to cover the possible range of the background fraction.
The background shape uncertainty is estimated by using the two 
background shapes, $F_1$ and $F_2$, while keeping the integrated
number of the background events in the signal region
( $P_T^2 < 0.1$ (GeV/c)$^2$) constant.

The shape of the soft-pion momentum distributions in 
$b \rightarrow D^*$ or $c \rightarrow D^*$
is determined by fitting to the MC distributions.
The uncertainty concerning this distribution 
is estimated by performing the analysis using 
a binned momentum distribution instead of fitting.

The total systematic errors are obtained to be 
$\pm$0.067 and $\pm$0.053 for 1993-95 and 1996-98, respectively.
\subsection{ Results }
The $A_c$ values obtained in the inclusive soft-pion analysis 
are
$A_c = 0.669 \pm 0.127(stat.) \pm 0.067(sys.)$ (1993-95) and 
$A_c = 0.689 \pm 0.057(stat.) \pm 0.053(sys.)$ (1996-98).
The combined result is
$$A_c = 0.685 \pm 0.052(stat.) \pm 0.038(sys.)$$

\section{ Conclusion }
Using the 1993-98 experimental data collected by the SLD experiment,
we measure the parity-violation parameter $A_c$ 
using two different $c$-quark tagging methods:
$$A_c = 0.690 \pm 0.042(stat.) \pm 0.019(sys.) \mbox{ and}$$
$$A_c = 0.685 \pm 0.052(stat.) \pm 0.036(sys.),$$
from exclusive charmed-meson reconstruction and inclusive soft-pion
analysis, respectively.

To combine them, we must avoid double counting signal events from both samples.
We find that 1182 events are common to the two analyses.
The statistical error for the soft-pion analysis without the
overlapping events is $\pm0.061$. 
The combined result is
$$A_c = 0.688 \pm 0.041,$$
where we have also treated the common systematic errors as fully correlated.

The result is consistent with the standard model prediction
of 0.667, obtained by using ZFITTER(6.23) 
with a top-quark mass of 175 GeV/c$^2$, 
and a Higgs mass of 150 GeV/c$^2$.
This result represents the currently most precise measurement of $A_c$.

\section*{Acknowledgments}
We thank the personnel of the SLAC accelerator department and the
technical
staffs of our collaborating institutions for their outstanding efforts
on our behalf.
This work was supported by the Department of Energy; 
the National Science Foundation; 
the Istituto Nazionale di Fisica Nucleare of Italy;
the Japan-US Cooperative Research Project on High Energy Physics;
and the Science and Engineering Research Council of the United
Kingdom.


\section*{$^{**}$List of Authors} 

%
%
%
\begin{center}

\def\iAOMORI{$^{(1)}$}
\def\iBRI{$^{(2)}$}
\def\iBRUN{$^{(3)}$}
\def\iBU{$^{(4)}$}
\def\iCOLO{$^{(5)}$}
\def\iCSU{$^{(6)}$}
\def\iFERR{$^{(7)}$}
\def\iFRAS{$^{(8)}$}
\def\iJHU{$^{(9)}$}
\def\iLBL{$^{(10)}$}
\def\iMASS{$^{(11)}$}
\def\iMISSI{$^{(12)}$}
\def\iMIT{$^{(13)}$}
\def\iMOSCOW{$^{(14)}$}
\def\iNAGO{$^{(15)}$}
\def\iOREG{$^{(16)}$}
\def\iOXF{$^{(17)}$}
\def\iPERU{$^{(18)}$}
\def\iRAL{$^{(19)}$}
\def\iRUTG{$^{(20)}$}
\def\iSLAC{$^{(21)}$}
\def\iSOONG{$^{(22)}$}
\def\iTENN{$^{(23)}$}
\def\iTOHO{$^{(24)}$}
\def\iUCSB{$^{(25)}$}
\def\iUCSC{$^{(26)}$}
\def\iVAND{$^{(27)}$}
\def\iWASH{$^{(28)}$}
\def\iWISC{$^{(29)}$}
\def\iYALE{$^{(30)}$}

\baselineskip=.75\baselineskip 

\mbox{Kenji Abe\unskip,\iNAGO}
\mbox{Koya Abe\unskip,\iTOHO}
\mbox{T. Abe\unskip,\iSLAC}
\mbox{I. Adam\unskip,\iSLAC}
\mbox{H. Akimoto\unskip,\iSLAC}
\mbox{D. Aston\unskip,\iSLAC}
\mbox{K.G. Baird\unskip,\iMASS}
\mbox{C. Baltay\unskip,\iYALE}
\mbox{H.R. Band\unskip,\iWISC}
\mbox{T.L. Barklow\unskip,\iSLAC}
\mbox{J.M. Bauer\unskip,\iMISSI}
\mbox{G. Bellodi\unskip,\iOXF}
\mbox{R. Berger\unskip,\iSLAC}
\mbox{G. Blaylock\unskip,\iMASS}
\mbox{J.R. Bogart\unskip,\iSLAC}
\mbox{G.R. Bower\unskip,\iSLAC}
\mbox{J.E. Brau\unskip,\iOREG}
\mbox{M. Breidenbach\unskip,\iSLAC}
\mbox{W.M. Bugg\unskip,\iTENN}
\mbox{D. Burke\unskip,\iSLAC}
\mbox{T.H. Burnett\unskip,\iWASH}
\mbox{P.N. Burrows\unskip,\iOXF}
\mbox{A. Calcaterra\unskip,\iFRAS}
\mbox{R. Cassell\unskip,\iSLAC}
\mbox{A. Chou\unskip,\iSLAC}
\mbox{H.O. Cohn\unskip,\iTENN}
\mbox{J.A. Coller\unskip,\iBU}
\mbox{M.R. Convery\unskip,\iSLAC}
\mbox{V. Cook\unskip,\iWASH}
\mbox{R.F. Cowan\unskip,\iMIT}
\mbox{G. Crawford\unskip,\iSLAC}
\mbox{C.J.S. Damerell\unskip,\iRAL}
\mbox{M. Daoudi\unskip,\iSLAC}
\mbox{S. Dasu\unskip,\iWISC}
\mbox{N. de Groot\unskip,\iBRI}
\mbox{R. de Sangro\unskip,\iFRAS}
\mbox{D.N. Dong\unskip,\iMIT}
\mbox{M. Doser\unskip,\iSLAC}
\mbox{R. Dubois\unskip,\iSLAC}
\mbox{I. Erofeeva\unskip,\iMOSCOW}
\mbox{V. Eschenburg\unskip,\iMISSI}
\mbox{E. Etzion\unskip,\iWISC}
\mbox{S. Fahey\unskip,\iCOLO}
\mbox{D. Falciai\unskip,\iFRAS}
\mbox{J.P. Fernandez\unskip,\iUCSC}
\mbox{K. Flood\unskip,\iMASS}
\mbox{R. Frey\unskip,\iOREG}
\mbox{E.L. Hart\unskip,\iTENN}
\mbox{K. Hasuko\unskip,\iTOHO}
\mbox{S.S. Hertzbach\unskip,\iMASS}
\mbox{M.E. Huffer\unskip,\iSLAC}
\mbox{X. Huynh\unskip,\iSLAC}
\mbox{M. Iwasaki\unskip,\iOREG}
\mbox{D.J. Jackson\unskip,\iRAL}
\mbox{P. Jacques\unskip,\iRUTG}
\mbox{J.A. Jaros\unskip,\iSLAC}
\mbox{Z.Y. Jiang\unskip,\iSLAC}
\mbox{A.S. Johnson\unskip,\iSLAC}
\mbox{J.R. Johnson\unskip,\iWISC}
\mbox{R. Kajikawa\unskip,\iNAGO}
\mbox{M. Kalelkar\unskip,\iRUTG}
\mbox{H.J. Kang\unskip,\iRUTG}
\mbox{R.R. Kofler\unskip,\iMASS}
\mbox{R.S. Kroeger\unskip,\iMISSI}
\mbox{M. Langston\unskip,\iOREG}
\mbox{D.W.G. Leith\unskip,\iSLAC}
\mbox{V. Lia\unskip,\iMIT}
\mbox{C. Lin\unskip,\iMASS}
\mbox{G. Mancinelli\unskip,\iRUTG}
\mbox{S. Manly\unskip,\iYALE}
\mbox{G. Mantovani\unskip,\iPERU}
\mbox{T.W. Markiewicz\unskip,\iSLAC}
\mbox{T. Maruyama\unskip,\iSLAC}
\mbox{A.K. McKemey\unskip,\iBRUN}
\mbox{R. Messner\unskip,\iSLAC}
\mbox{K.C. Moffeit\unskip,\iSLAC}
\mbox{T.B. Moore\unskip,\iYALE}
\mbox{M. Morii\unskip,\iSLAC}
\mbox{D. Muller\unskip,\iSLAC}
\mbox{V. Murzin\unskip,\iMOSCOW}
\mbox{S. Narita\unskip,\iTOHO}
\mbox{U. Nauenberg\unskip,\iCOLO}
\mbox{H. Neal\unskip,\iYALE}
\mbox{G. Nesom\unskip,\iOXF}
\mbox{N. Oishi\unskip,\iNAGO}
\mbox{D. Onoprienko\unskip,\iTENN}
\mbox{L.S. Osborne\unskip,\iMIT}
\mbox{R.S. Panvini\unskip,\iVAND}
\mbox{C.H. Park\unskip,\iSOONG}
\mbox{I. Peruzzi\unskip,\iFRAS}
\mbox{M. Piccolo\unskip,\iFRAS}
\mbox{L. Piemontese\unskip,\iFERR}
\mbox{R.J. Plano\unskip,\iRUTG}
\mbox{R. Prepost\unskip,\iWISC}
\mbox{C.Y. Prescott\unskip,\iSLAC}
\mbox{B.N. Ratcliff\unskip,\iSLAC}
\mbox{J. Reidy\unskip,\iMISSI}
\mbox{P.L. Reinertsen\unskip,\iUCSC}
\mbox{L.S. Rochester\unskip,\iSLAC}
\mbox{P.C. Rowson\unskip,\iSLAC}
\mbox{J.J. Russell\unskip,\iSLAC}
\mbox{O.H. Saxton\unskip,\iSLAC}
\mbox{T. Schalk\unskip,\iUCSC}
\mbox{B.A. Schumm\unskip,\iUCSC}
\mbox{J. Schwiening\unskip,\iSLAC}
\mbox{V.V. Serbo\unskip,\iSLAC}
\mbox{G. Shapiro\unskip,\iLBL}
\mbox{N.B. Sinev\unskip,\iOREG}
\mbox{J.A. Snyder\unskip,\iYALE}
\mbox{H. Staengle\unskip,\iCSU}
\mbox{A. Stahl\unskip,\iSLAC}
\mbox{P. Stamer\unskip,\iRUTG}
\mbox{H. Steiner\unskip,\iLBL}
\mbox{D. Su\unskip,\iSLAC}
\mbox{F. Suekane\unskip,\iTOHO}
\mbox{A. Sugiyama\unskip,\iNAGO}
\mbox{S. Suzuki\unskip,\iNAGO}
\mbox{M. Swartz\unskip,\iJHU}
\mbox{F.E. Taylor\unskip,\iMIT}
\mbox{J. Thom\unskip,\iSLAC}
\mbox{E. Torrence\unskip,\iMIT}
\mbox{T. Usher\unskip,\iSLAC}
\mbox{J. Va'vra\unskip,\iSLAC}
\mbox{R. Verdier\unskip,\iMIT}
\mbox{D.L. Wagner\unskip,\iCOLO}
\mbox{A.P. Waite\unskip,\iSLAC}
\mbox{S. Walston\unskip,\iOREG}
\mbox{A.W. Weidemann\unskip,\iTENN}
\mbox{E.R. Weiss\unskip,\iWASH}
\mbox{J.S. Whitaker\unskip,\iBU}
\mbox{S.H. Williams\unskip,\iSLAC}
\mbox{S. Willocq\unskip,\iMASS}
\mbox{R.J. Wilson\unskip,\iCSU}
\mbox{W.J. Wisniewski\unskip,\iSLAC}
\mbox{J.L. Wittlin\unskip,\iMASS}
\mbox{M. Woods\unskip,\iSLAC}
\mbox{T.R. Wright\unskip,\iWISC}
\mbox{R.K. Yamamoto\unskip,\iMIT}
\mbox{J. Yashima\unskip,\iTOHO}
\mbox{S.J. Yellin\unskip,\iUCSB}
\mbox{C.C. Young\unskip,\iSLAC}
\mbox{H. Yuta\unskip.\iAOMORI}

\it
\vskip \baselineskip                   
\vskip \baselineskip
\baselineskip=.75\baselineskip   
\iAOMORI
  Aomori University, Aomori , 030 Japan, \break
\iBRI
  University of Bristol, Bristol, United Kingdom, \break
\iBRUN
  Brunel University, Uxbridge, Middlesex, UB8 3PH United Kingdom, \break
\iBU
  Boston University, Boston, Massachusetts 02215, \break
\iCOLO
  University of Colorado, Boulder, Colorado 80309, \break
\iCSU
  Colorado State University, Ft. Collins, Colorado 80523, \break
\iFERR
  INFN Sezione di Ferrara and Universita di Ferrara, I-44100 Ferrara, Italy, \break
\iFRAS
  INFN Lab. Nazionali di Frascati, I-00044 Frascati, Italy, \break
\iJHU
  Johns Hopkins University,  Baltimore, Maryland 21218-2686, \break
\iLBL
  Lawrence Berkeley Laboratory, University of California, Berkeley, California 94720, \break
\iMASS
  University of Massachusetts, Amherst, Massachusetts 01003, \break
\iMISSI
  University of Mississippi, University, Mississippi 38677, \break
\iMIT
  Massachusetts Institute of Technology, Cambridge, Massachusetts 02139, \break
\iMOSCOW
  Institute of Nuclear Physics, Moscow State University, 119899, Moscow Russia, \break
\iNAGO
  Nagoya University, Chikusa-ku, Nagoya, 464 Japan, \break
\iOREG
  University of Oregon, Eugene, Oregon 97403, \break
\iOXF
  Oxford University, Oxford, OX1 3RH, United Kingdom, \break
\iPERU
  INFN Sezione di Perugia and Universita di Perugia, I-06100 Perugia, Italy, \break
\iRAL
  Rutherford Appleton Laboratory, Chilton, Didcot, Oxon OX11 0QX United Kingdom, \break
\iRUTG
  Rutgers University, Piscataway, New Jersey 08855, \break
\iSLAC
  Stanford Linear Accelerator Center, Stanford University, Stanford, California 94309, \break
\iSOONG
  Soongsil University, Seoul, Korea 156-743, \break
\iTENN
  University of Tennessee, Knoxville, Tennessee 37996, \break
\iTOHO
  Tohoku University, Sendai 980, Japan, \break
\iUCSB
  University of California at Santa Barbara, Santa Barbara, California 93106, \break
\iUCSC
  University of California at Santa Cruz, Santa Cruz, California 95064, \break
\iVAND
  Vanderbilt University, Nashville,Tennessee 37235, \break
\iWASH
  University of Washington, Seattle, Washington 98105, \break
\iWISC
  University of Wisconsin, Madison,Wisconsin 53706, \break
\iYALE
  Yale University, New Haven, Connecticut 06511. \break

\rm
%

\end{center}

%
%


\begin{figure}
\begin{center}
\epsfysize 16cm
\epsfbox{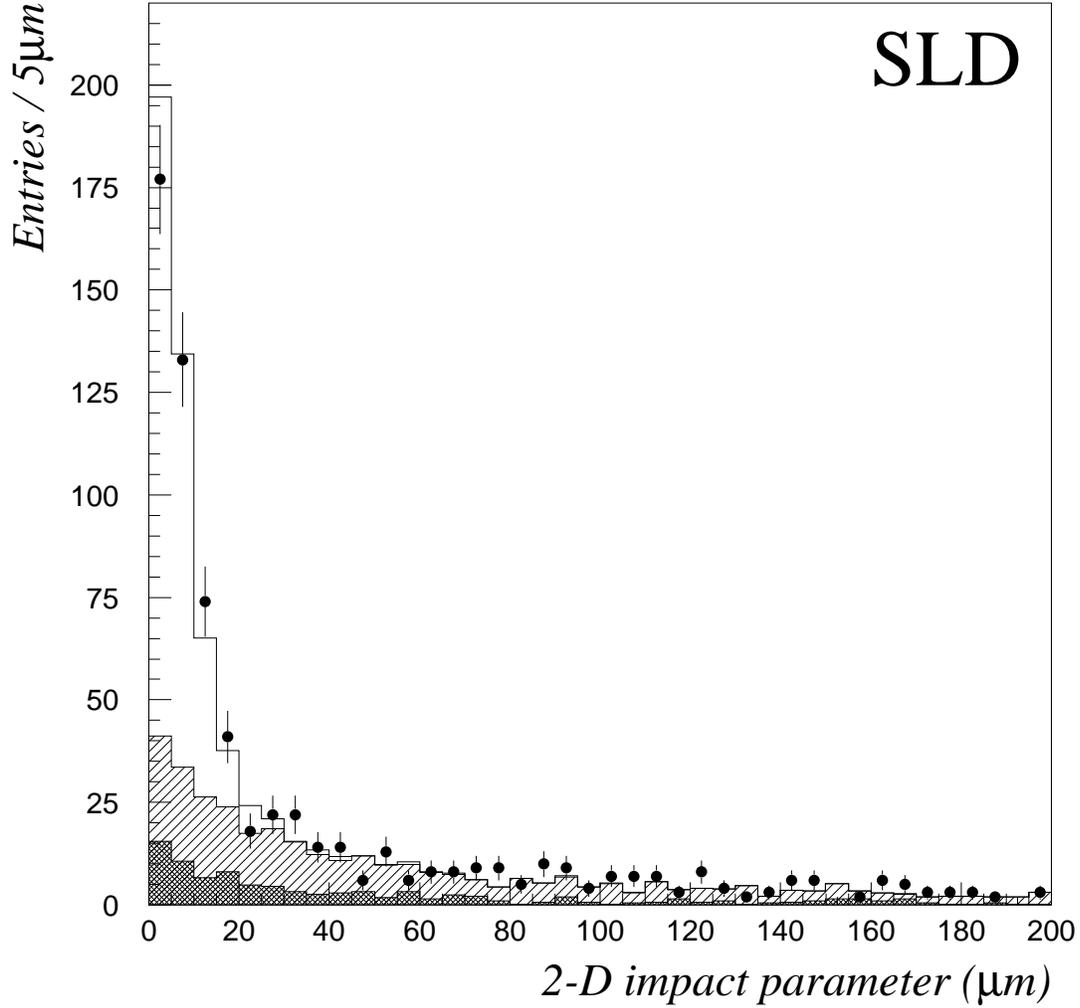}
\end{center}
\caption{
The distribution of the 2-D impact parameter 
of the $D^0$ momentum vector to the IP
for the decay of
$D^{\ast+}\rightarrow D^0 \pi_s^+$, $D^0 \rightarrow K^- \pi^+$.
The solid circles indicate the experimental data, and histograms are
MC of $D^{\ast+}$ from $c$-quark (open), from $b$-quark (single
hatched), and RCBG (double hatched).
}
\label{fig1}
\end{figure} 

\begin{figure}
\begin{center}
\epsfysize 16cm
\epsfbox{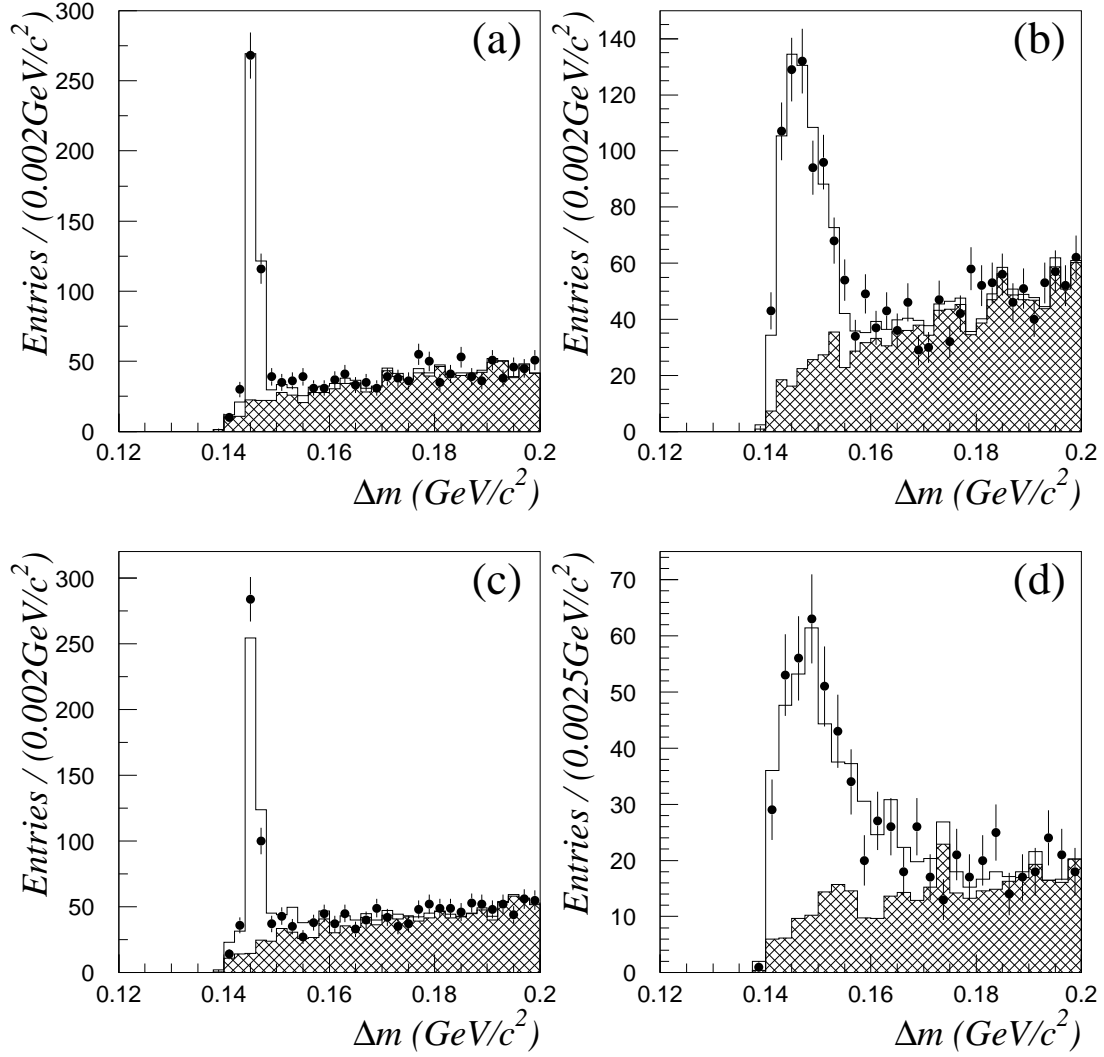}
\end{center}
\caption{
The mass-difference distributions for the decay of
(a)$D^{\ast+}\rightarrow D^0 \pi_s^+$, $D^0 \rightarrow K^- \pi^+$, 
(b)$D^0 \rightarrow K^- \pi^+ \pi^0$, 
(c)$D^0 \rightarrow K^- \pi^+ \pi^+ \pi^-$, and  
(d)$D^0 \rightarrow K^- l^+ \nu_l$ ($l=$e or $\mu$).
The solid circles indicate the experimental data, and histograms are
MC of signal (open) and RCBG (double hatched).
}
\label{fig2}
\end{figure} 

\begin{figure}
\begin{center}
\epsfysize 9cm
\epsfbox{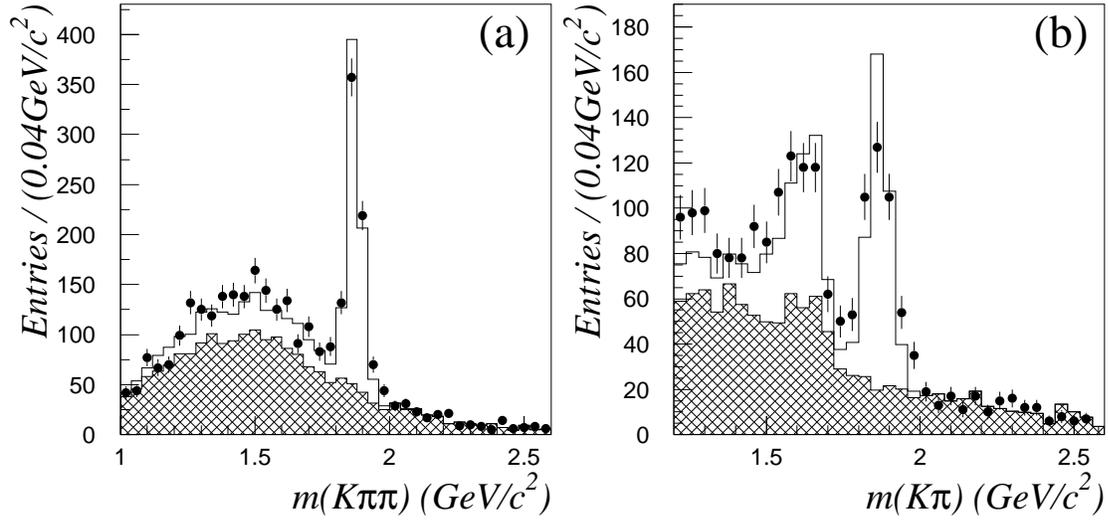}
\end{center}
\caption{
The mass distributions for (a)$D^+$ and (b)$D^0$ mesons.
The solid circles indicate the experimental data, and histograms are
the MC of signal (open) and RCBG (double hatched). 
The peaks around $m(K\pi) \sim$ 1.6 GeV/c$^2$ in figure (b) comes
from the decay $D^0 \rightarrow K \pi \pi^0$.
}
\label{fig3}
\end{figure} 

\begin{figure}
\begin{center}
\epsfysize16cm
\epsfbox{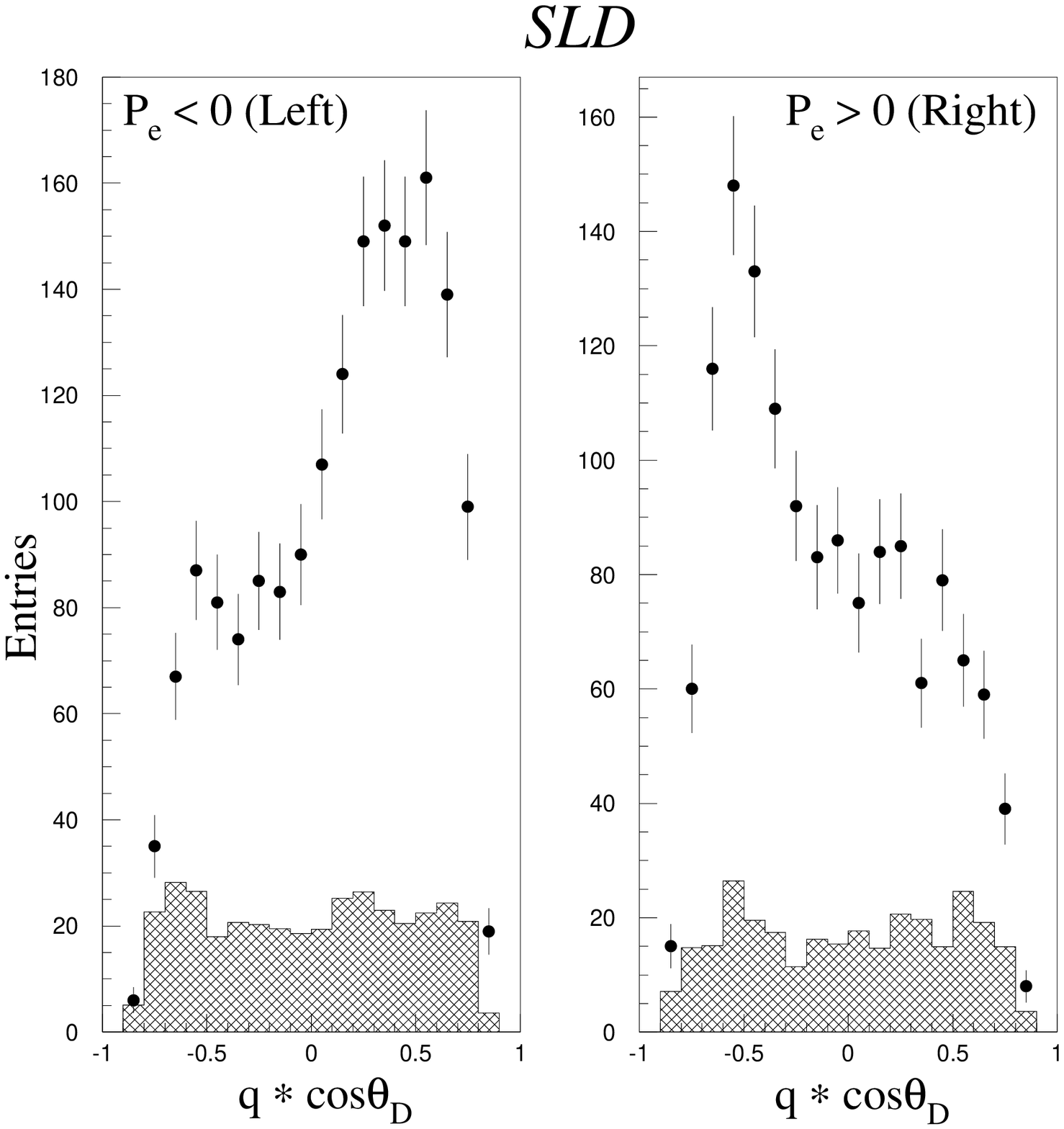}
\end{center}
\caption{
The distributions of $q \cdot \cos\theta_D$ for
the selected $D$ meson sample for (a) left- and (b) right-
handed electron beams.
The solid circles are experimental data, and double hatched histograms 
are RCBG estimated from side-band regions.
}
\label{fig4}
\end{figure} 

\begin{figure}
\begin{center}
\epsfysize16cm
\epsfbox{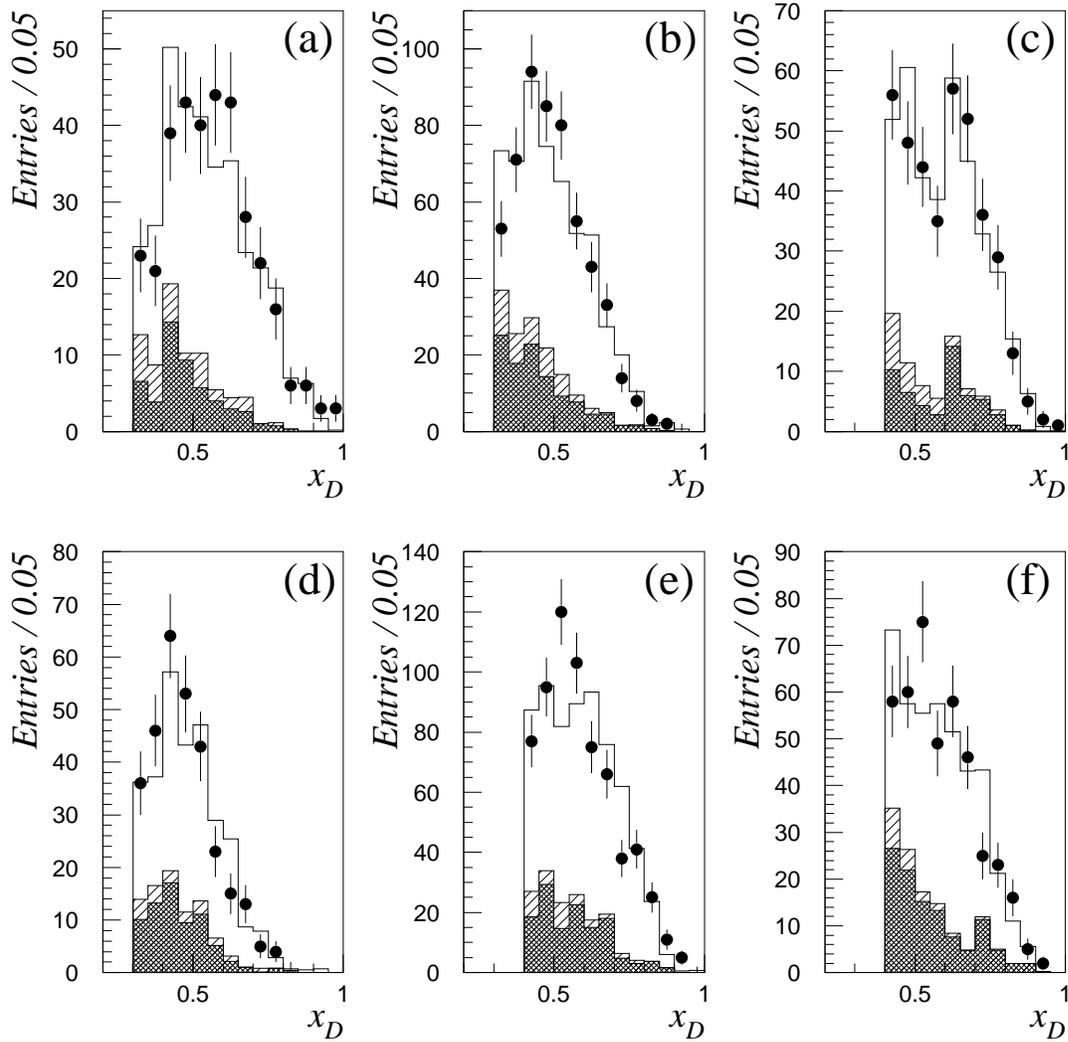}
\end{center}
\caption{
The $x_D$ distributions for 
(a)$D^{\ast+}\rightarrow D^0 \pi_s^+$, $D^0 \rightarrow K^- \pi^+$, 
(b)$D^0 \rightarrow K^- \pi^+ \pi^0$, 
(c)$D^0 \rightarrow K^- \pi^+ \pi^+ \pi^-$,  
(d)$D^0 \rightarrow K^- l^+ \nu_l$ ($l=$e or $\mu$)
(e)$D^+ \rightarrow K^- \pi^+ \pi^+$, and
(f)$D^0 \rightarrow K^- \pi^+$.
The solid circles are experimental data and 
hatched histograms are background estimated 
from side-band events. MC predictions for 
$D$ mesons from $c$-decay (open histograms) and 
$b$-decay (single hatched histogram) are also shown.
}
\label{fig5}
\end{figure} 

\begin{figure}
\begin{center}
\epsfysize9cm
\epsfbox{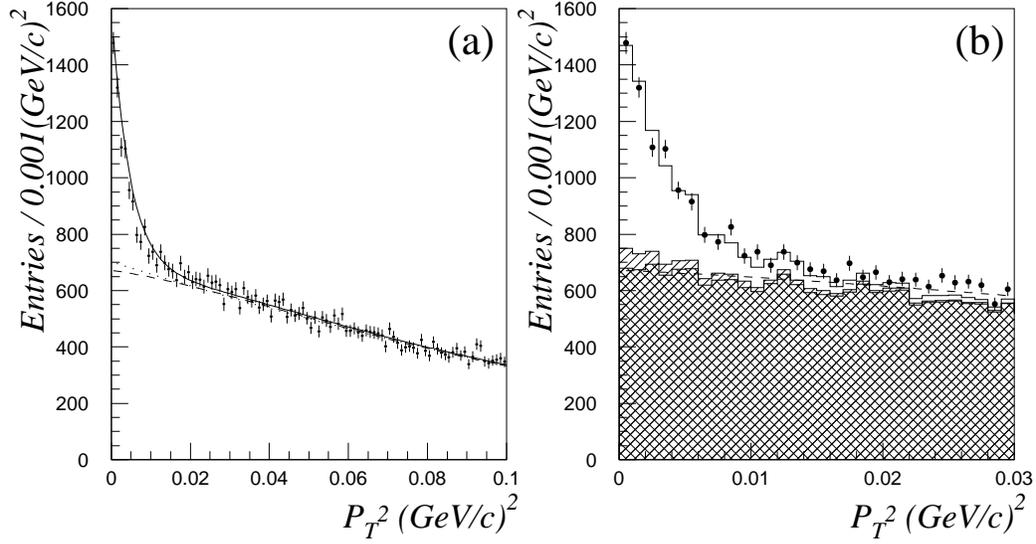}
\end{center}
\caption{
The $P_T^2$ distributions for soft-pion candidate tracks.
(a) The solid circles indicate the experimental data. 
The curves are the result of the a fit $S(P_T^2) + F_1(P_T^2)$
performed for $P_T^2 < 0.1$ GeV/c (solid line),
and the extrapolations of $F_1(P_T^2)$ (dashed line) and
$F_2(P_T^2)$ (dotted line).
The definition of the functions are described in the text.
(b) The solid circles are the experimental data, and histograms are
MC predictions for
$D$ mesons from $c$-decay (open),
$D$ mesons from $b$-decay (single hatched), 
and background (double hatched).
The extrapolation of $F_1(P_T^2)$ is also shown as a dashed line.
}
\label{fig6}
\end{figure} 

\begin{figure}
\begin{center}
\epsfysize16cm
\epsfbox{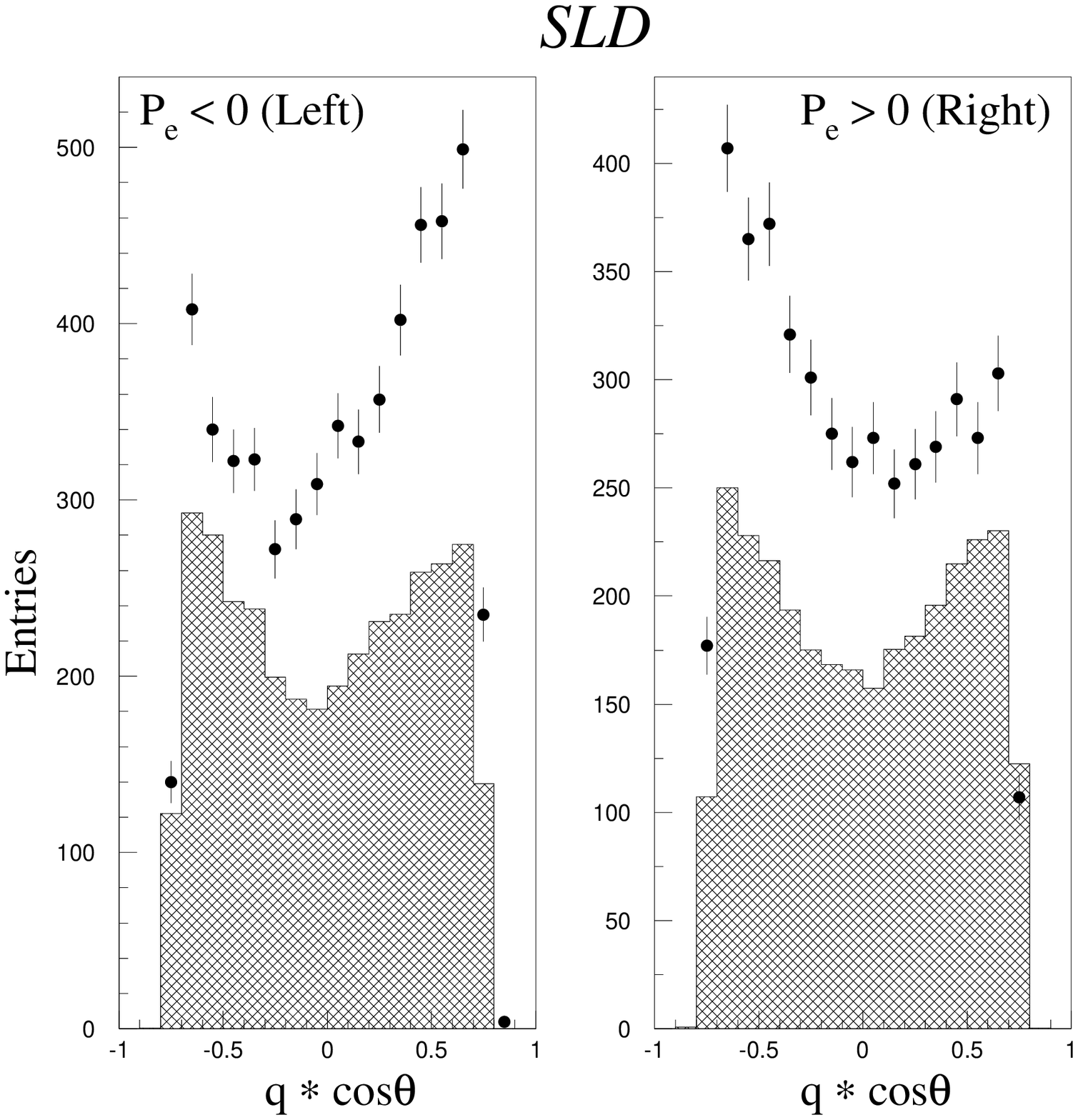}
\end{center}
\caption{
The distributions of $q \cdot \cos\theta_D$ for
the selected $D^{\ast +}$ meson sample for (a) left- and (b) right-
handed electron beams.
The solid circles are experimental data, and hatched histograms are RCBG
estimated from side-band regions.
}
\label{fig7}
\end{figure} 

\begin{figure}
\begin{center}
\epsfysize9cm
\hskip0.1in
\epsfbox{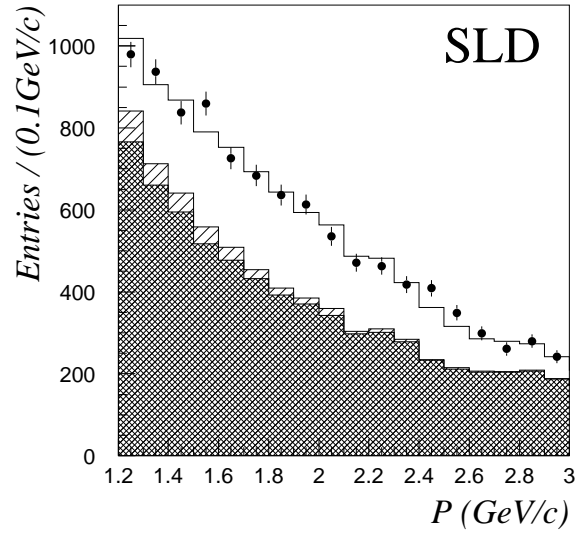}
\end{center}
\caption{
The momentum distribution for soft-pion candidate tracks. 
The points are experimental data. The histograms are
MC predictions of $D$'s from $c$ decays (open), 
$D$'s from $b$ decays (single hatched), and
background (double-hatched).
}
\label{fig8}
\end{figure} 


%
%

\begin{table}[h]
\caption{
The number of selected candidates from 1993-98 SLD experimental data,
and contributions from $c\rightarrow D$, $b\rightarrow D$, and RCBG
estimated by MC.
}
\label{tab:candidate}
\begin{center}
\begin{tabular}{l c r r r } 
\multicolumn{1}{l}{Channel} & 
\multicolumn{1}{c}{Candidates} &
\multicolumn{1}{c}{$c \rightarrow D$} &
\multicolumn{1}{c}{$b \rightarrow D$} &
\multicolumn{1}{c}{RCBG} 
\\
 \hline
$D^{\ast +}\rightarrow D^0 \pi_s^+$, & & & \\
\ \ \ 
$D^0 \rightarrow K^- \pi^+ $ & 
\ 561 & 413 (74\%)& 59 (10\%)&  89 (16\%)\\
\ \ \ 
$D^0 \rightarrow K^- \pi^+ \pi^0$ &
\ 896 & 601 (67\%)& 83 (\ 9\%)& 212 (24\%) \\
\ \ \ 
$D^0 \rightarrow K^- \pi^+ \pi^+ \pi^- $ &
\ 537 & 418 (78\%)& 36 (\ 7\%)&  83 (15\%)\\
\ \ \ 
$D^0 \rightarrow K^- l^+ \bar{\nu} $ &
\ 433 & 296 (68\%)& 31 (\ 7\%)& 106 (24\%)\\
 \hline
$D^+ \rightarrow K^- \pi^+ \pi^+ $ &
\ 957 & 698 (73\%)& 45 (\ 5\%)& 214 (22\%)\\
$D^0 \rightarrow K^- \pi^+ $ &
\ 583 & 403 (69\%)& 27 (\ 5\%)& 153 (26\%)\\
\hline
Total &
3967 & 2829 (71\%)& 281 (\ 7\%)& 857 (22\%)
\end{tabular}
\end{center}
\end{table}

\begin{table}[h]
\caption{
1993-98 average contributions to the estimated systematic error 
for exclusive $D$ meson reconstruction analysis (left column) and 
inclusive soft-pion analysis (right column).
}
\label{tab:systematic_ex}
\begin{center}
\begin{tabular}{l c c}
 & \multicolumn{2}{c}{$\delta A_c$} \\
\multicolumn{1}{l}{Source} 
& \multicolumn{1}{c}{Exclusive $D^{(\ast)}$}
& \multicolumn{1}{c}{Inclusive soft pion} \\
\hline 
Background fraction            & 0.0111 & 0.0324\\
Background acceptance          & 0.0087 & 0.0122\\
Background $x_D$ / $P_T^2$ distribution  
                               & 0.0112 & 0.0018\\
Background asymmetry           & 0.0028 & 0.0093\\
$f_{b \rightarrow D}/(f_{b \rightarrow D}+f_{c \rightarrow D})$ 
                               & 0.0011 & 0.0018\\
$A_{b \rightarrow D}$ $(A_b)$  & 0.0017 & 0.0021\\
$A_{b \rightarrow D}$ (Mixing) & 0.0092 & 0.0120\\
$c$ fragmentation              & 0.0003 & 0.0010\\
$b$ fragmentation              & 0.0003 & 0.0005\\
$D$ meson $x_D$ shape / Soft-pion momentum shape         
                               & 0.0040 & 0.0003\\
Polarization                   & 0.0035 & 0.0033\\
$A_e$                          & 0.0002 & 0.0005\\
$\alpha_s$                     & 0.0004 & 0.0005\\
Correction factor for first order QCD correction        
                               & 0.0024 & 0.0033\\
Second order QCD correction    & 0.0006 & 0.0008\\
Gluon splitting                & 0.0002 & 0.0005\\
\hline
Total                          & 0.0213 & 0.0383
\end{tabular}
\end{center}
\end{table}

\end{document}